\documentstyle[preprint,aps]{revtex}

\begin{document}

\draft

\title{ Bosons in anisotropic traps:\\
ground state and vortices}

\author{F. Dalfovo and S. Stringari}

\address{Dipartimento di Fisica, Universit\`a di Trento, \\
        and Istituto Nazionale di Fisica della Materia, \\
        I-38050 Povo,  Italy }

\date{ 25 October 1995 }
\maketitle

\begin{abstract}
We solve the Gross-Pitaevskii equations for a dilute atomic gas in
a magnetic trap, modeled by an anisotropic harmonic potential. We
evaluate the wave function and the  energy of the Bose Einstein
condensate as a function  of the particle number, both for  positive
and negative scattering length. The results for the transverse and
vertical size of the cloud of atoms, as well as for the kinetic and
potential energy per particle, are compared with the predictions of
approximated models. We also compare the aspect ratio of the velocity
distribution  with first experimental estimates available for
$^{87}$Rb.  Vortex states are considered and the critical angular
velocity for production of vortices is calculated. We show that the
presence of vortices  significantly increases the stability of the
condensate in the case of attractive interactions.
\end{abstract}


\narrowtext

\section{Introduction}
\label{sec:intro}

Recent experiments support the existence of a Bose Einstein condensate
in vapors of alkali atoms, such as rubidium\cite{And95},
litium\cite{Bra95} and sodium\cite{Dav95},  confined in magnetic traps
and  cooled down to temperature of the order of $100$ nanokelvin. This
important discovery opens new interesting perspectives in the field of
many-body physics (for a comprehensive review on Bose Einstein
condensation see for instance Ref.~\cite{Gri95}).

The  vapors of alkali atoms used in the experiments are very dilute,
i.e., the average  distance among the atoms is much larger than the
range of the interaction. So the physics  is expected to be dominated
by   two-body collisions, well accounted for by the
knowledge  of the $s$-wave scattering length. This also implies  that
the Gross-Pitaevskii theory\cite{Pit61} for weakly interacting bosons
finds  in these systems an ideal field of application.
This  theory has been already used to describe bosons confined in
isotropic traps\cite{Edw95,Rup95,Fet95}. The anisotropic case has been
recently discussed by Baym and Pethick\cite{Bay95}, who have obtained
approximate analytic solutions, thereby providing a first qualitative
insight into many interesting features  of these systems. To make the
analysis more quantitative one has to solve numerically the
Gross-Pitaevskii equations.

In  the present work we provide a complete set of solutions  of the
Gross-Pitaevskii equations  applied to $N$ alkali atoms in anisotropic
traps. We calculate the condensate wave function at $T=0$ for bosons
interacting  through positive and negative scattering length. We
explore the $N$-dependence of  relevant quantities,  such as energy,
chemical potential and the aspect ratio of the velocity
distribution.  We also discuss vortex states, studying the density
profile and the critical angular velocity.

The results  of the numerical minimization of the  energy functional
are compared with the analytic solution of the noninteracting
anisotropic harmonic oscillator as well as with approximated models
available when the interaction is strongly repulsive (strongly repulsive
limit).   For instance we will show that the surface structure of the
condensate wave function,  for which a sound treatment of the kinetic
energy is needed, is relevant in determining the aspect ratio. Our
results for the aspect ratio in $^{87}$Rb, with  $N$ of the order of
a few thousands,  are in agreement with the first experimental
findings of Ref.~\cite{And95}. An accurate determination of the
particle distribution in the vapor cloud is also relevant in order to
calculate  the moment of inertia of the system, which is directly
connected with the superfluid  behavior\cite{Str95}.

We find  instabilities in the solutions  for negative scattering
length when the number of atoms exceeds certain critical values, of
the order of $1400$ for $^7$Li, in agreement with previous
analysis\cite{Rup95}. However we find also that the stability of
these systems is significantly increased when vortex states are
considered. This happens because  the vortex flow pushes the atoms
away from the center of the trap decreasing the highest value of the
local density.  For $^7$Li we find minima of the Gross-Pitaevskii
functional corresponding to vortex states with quantum of circulation
higher than $1$ and  having  $N$   of the order of $10000$.

The paper is organized as follows. In Sec.~\ref{section2} we present
the formalism of  the Gross-Pitaevskii theory for
anisotropic traps and we discuss the solutions in two limits
(noninteracting and strongly repulsive limit).  We also
discuss the generalization of the theory to include vortex states. In
Sec.~\ref{section3}  we briefly introduce the numerical procedure,
which is based on the steepest descent method for functional
minimization. In Sec.~\ref{section4} we present the results for the
two cases of  positive ($^{87}$Rb) and negative ($^7$Li)  scattering
length. Finally, in Sec.~\ref{conclusion} we   summarize the main
results.

\section{Gross-Pitaevskii theory for trapped bosons}
\label{section2}

In the Gross-Pitaevskii theory\cite{Pit61} the ground state energy for
condensed  bosons of mass $m$ is given by the  functional
\begin{equation}
E[\psi] = \int\!d{\bf r} \left[
{\hbar^2 \over 2m} |\nabla \psi({\bf r})|^2
+ {m\over 2} (\omega_\perp^2 x^2  + \omega_\perp^2 y^2 +
\omega_z^2  z^2) |\psi({\bf r})|^2 + {2\pi\hbar^2 a \over m}
|\psi({\bf r})|^4 \right] \; ,
\label{eq:e}
\end{equation}
where $\psi({\bf r})$ is the condensate wave function (order
parameter),   $\omega_\perp$  and $\omega_z$ are the two angular
frequencies  associated with the  external  potential of the
anisotropic trap and $a$ is the $s$-wave scattering length.  The wave
function is assumed to be normalized to the number of atoms in the
condensate:
\begin{equation}
\int\! d{\bf r} \ |\psi({\bf r})|^2  = N \; .
\label{eq:normpsi}
\end{equation}
In the  $T \to 0$ limit considered in this work,  $N$ coincides  with
the total  number of atoms in  the trap. The explicit form of the
ground state  wave function is obtained  by minimizing the energy
functional.  One can also write explicitly the  variation of the
energy functional at first order, finding the  Euler-Lagrange equation
\begin{equation}
\left[ - {\hbar^2 \over 2m} \nabla^2 + {m \over 2} (\omega_\perp^2 x^2
+ \omega_\perp^2 y^2 +\omega_z^2 z^2) + {4\pi \hbar^2 a \over m}
|\psi({\bf r})|^2 \right] \psi({\bf r}) = \mu \psi({\bf r}) \; ,
\label{eq:nlse}
\end{equation}
where $\mu$  is the chemical potential. This equation  has the form
of a  nonlinear stationary Schr\"odinger equation. It has been
recently  solved  for bosons in isotropic harmonic traps by Edwards
and Burnett\cite{Edw95},  by means of iterated Runge-Kutta
integrations. An efficient numerical  calculation of the ground
state, which works well also in the case of  anisotropic traps with
and without vortices, consists in minimizing  directly  the energy
functional (\ref{eq:e}) with a steepest  descent  method, as we will
show in sect.~\ref{section3}.

The Gross-Pitaevskii theory is expected to be accurate when  the
system is  dilute. If $\rho$ is the density of bosons, the  parameter
which measures the  applicability of the theory is the product $a^3
\rho$, which should be much less then $1$. This condition is largely
satisfied by the samples of alkali atoms used in the recent
experiments\cite{And95,Bra95}. For instance, the central density of
$10000$ atoms of $^{87}$Rb in the trap of  Ref.~\cite{And95} is
expected to be of the order of $10^{12} \div 10^{13}$ cm$^{-3}$, which
yields $a^3 \rho \simeq 10^{-6}$ or less. Even in the experiment of
Ref.~\cite{Dav95} on sodium, where the number of atoms in the
condensate is much larger, the quantity $a^3 \rho$ remains very
small ($\simeq 10^{-5}$).  The theory  is also accurate for dilute
systems of bosons having negative scattering  length. In this case,
however, one has to take care of the possible  instability induced by
the attractive interaction when $N$ is large.  When the  system
collapses, a more accurate theory is required in order to include
short range effects.

To simplify  the formalism one can choose a slightly different
notation,  taking  advantage of the fact that all distances and
energies in  the calculation scale as the typical length and energy
of the harmonic  external potential. So, we  introduce the standard
lengths
\begin{equation}
a_\perp = \left( {\hbar \over m \omega_\perp} \right)^{1/2} \; ; \; \;
a_z = \left( {\hbar \over m \omega_z} \right)^{1/2}  \; ,
\label{eq:aperpaz}
\end{equation}
and we rescale the spatial coordinate, the energy and the wave
function as follows:
\begin{equation}
{\bf r} = a_\perp {\bf r}_1
\label{eq:r1}
\end{equation}
\begin{equation}
E = \hbar \omega_\perp E_1
\label{eq:e1}
\end{equation}
\begin{equation}
\psi ({\bf r}) = \sqrt{N/a_\perp^3} \ \psi_1 ({\bf r}_1) \; .
\label{eq:psi1}
\end{equation}
The wave function $\psi_1$ is normalized to $1$.
Finally we introduce the asymmetry parameter
\begin{equation}
\lambda = \omega_z /  \omega_\perp
\label{eq:lambda}
\end{equation}
and define the quantity
\begin{equation}
u_1 = { 8 \pi a N \over a_\perp } \; .
\label{eq:u1}
\end{equation}
With these changes, the Gross-Pitaevskii energy functional  takes
the form
\begin{equation}
{ E_1 \over N } = \int\! d{\bf r}_1  \left[ |\nabla_1 \psi_1 ({\bf
r}_1)|^2 + (x_1^2 + y_1^2 + \lambda^2 z_1^2) |\psi_1 ({\bf r}_1)|^2 +
{u_1 \over 2}  |\psi_1 ({\bf r}_1)|^4 \right]
\label{eq:functional}
\end{equation}
and the nonlinear Schr\"odinger equation becomes
\begin{equation}
\left[ -\nabla_1^2 + x_1^2 + y_1^2  + \lambda^2 z_1^2 +
u_1 |\psi_1 ({\bf r}_1)|^2 \right] \psi_1 ({\bf r}_1) =
2 \mu_1 \psi_1 ({\bf r}_1)  \; ,
\label{eq:nlse1}
\end{equation}
where $\mu=\hbar \omega_\perp \mu_1$. The  dimensionless quantity
$u_1$ characterizes the effect of the interaction in the equation
for the condensate.  It is worth noting that even  if the system is
very dilute ($a^3 \rho \ll 1$) the interactions can nevertheless play
a crucial role in determining the solution of the Gross-Pitaevskii
equations. In fact the two conditions $a^3 \rho \ll 1$ and $u_1
\simeq 1$ can be well satisfied for realistic values of the
parameters of the problem, i.e., the scattering length, the
oscillator frequencies and the number of atoms in the trap. We now
discuss briefly the two limiting cases of noninteracting  particles
and of strongly repulsive interactions, where the solution of
Eq.(\ref{eq:nlse1}) is available in analytic way. We then show how
the formalism can be extended to describe  vortex states.

\subsection{The noninteracting model}
\label{subsectiona}

When the scattering length $a$ vanishes,  the problem reduces to the
solution of the stationary Schr\"odinger equation for an anisotropic
harmonic oscillator. The ground state wave function is
\begin{equation}
\psi_1 ({\bf r}_1) = \lambda^{1/4} \pi^{-3/4} \exp  \left[ - {1 \over
2}  ( x_1^2 + y_1^2 + \lambda z_1^2) \right] \; .
\label{eq:psi1ho}
\end{equation}
The chemical potential coincides with the energy per particle, which
turns  out to be $(1+\lambda/2)$. The gaussian has different
transverse and vertical widths.  In particular one has $\langle x_1^2
\rangle = \langle y_1^2 \rangle  = 1/2$ and $\langle z_1^2 \rangle =
1/(2\lambda)$. The mean values of the momentum operators $p_x^2$ and
$p_z^2$ can be  also easily  calculated. In particular it is
interesting to calculate the quantity
\begin{equation}
\sqrt{ \langle p^2_z \rangle /  \langle p^2_x \rangle }
=  \sqrt{ \langle x_1^2 \rangle /
\langle z_1^2 \rangle } =
\sqrt{\lambda}
\label{eq:arho}
\end{equation}
In the following we consider the quantity  $\sqrt{ \langle
p^2_z \rangle /  \langle p^2_x \rangle }$ as a measure of
the {\it aspect ratio} which characterizes  the anisotropy of the
velocity distribution. This is  a relevant quantity in the
interpretation of the experimental results. Values of the aspect ratio
different from $1$ reflect a peculiar and unique feature of  Bose
Einstein condensation.

\subsection{Strongly repulsive limit }
\label{subsectionb}

The opposite limit is obtained when the interaction is so strong, or
the number of particles so large, that the kinetic energy can be
neglected  in the energy functional. It corresponds to very large
values of the parameter $u_1$ (see Eq.~(\ref{eq:u1})). This limit has
been already discussed in  Refs.~\cite{Edw95,Bay95}. The solution is
easily obtained by dropping  the kinetic energy term in
Eq.~(\ref{eq:nlse1}). It has the form   \begin{equation}
\psi_1^2 ({\bf r}_1) = {1 \over u_1} ( 2 \mu_1 - x_1^2 - y_1^2 -
\lambda^2 z_1^2)
\label{eq:thomasfermi}
\end{equation}
if the right hand side is positive, and $\psi_1=0$ elsewhere. The
chemical potential is easily  calculated by imposing the
normalization condition $\int \psi_1^2 d{\bf r}_1=1$. One  finds
\begin{equation}
2 \mu_1 = \left( {15 \over 8
\pi} \lambda u_1 \right)^{2/5} \; ,
\label{eq:mu1}
\end{equation}
and $\mu=\hbar \omega_\perp \mu_1$. Using the definition of  $u_1$
given in  Eq.~(\ref{eq:u1})  and the relation $\mu=dE/dN$,  one also
gets the relationship $E/N=(5/7)\mu$.  The cloud of atoms extends
over a radius $R_1 = \sqrt{ 2\mu_1}$ and a vertical size
$Z_1=\lambda R_1$. One also finds the results
\begin{equation}
\langle x_1^2 \rangle = {2 \mu_1 \over 7}  \;  ; \;
\langle z_1^2 \rangle =  {2 \mu_1 \over 7 \lambda^2}
\label{eq:z1ave}
\end{equation}
for the square radius $\langle x_1^2 \rangle$ and $\langle z_1^2
\rangle$.  Due to the different scaling properties of the wave
function with respect to the variable $z$ (compare
Eqs.~(\ref{eq:psi1ho}) and (\ref{eq:thomasfermi})), the aspect ratio
$\sqrt{ \langle p^2_z \rangle /   \langle p^2_x \rangle }$ in this
case is equal to $\lambda$ and not to $\sqrt{\lambda}$ as in the
noninteracting case.   The central density of the cloud is   $\sqrt{2
\mu_1 / u_1}$.  The wave function (\ref{eq:thomasfermi}) is expected
to approximate well  the  exact solution of the nonlinear
Schr\"odinger equation (\ref{eq:nlse1}) for  large  $N$, apart from
the structure of the surface region where the exact  wave  function
has to vanish smoothly. This fact has been already tested  for
isotropic traps in Ref.~\cite{Edw95}. However some relevant
observables can be   significantly affected by this surface structure
even at large  $N$, as we will see later.

\subsection{Vortex states}
\label{subsectionc}

The energy functional (\ref{eq:functional}) is easily generalized to
include vortex states, i.e., states where all the atoms rotate around
the $z$-axis with quantized circulation. Indeed one of the primary
motivations of the Gross-Pitaevskii theory was the study of
vortex states in weakly interacting bosons\cite{Pit61}. One has to
consider a complex condensate wave function of the form
\begin{equation}
\Psi ({\bf r}) = \psi ({\bf r}) \exp [i S({\bf r})]
\label{eq:psi1expis}
\end{equation}
where $\psi({\bf r})= \sqrt{\rho({\bf r})}$ is the modulus, while the
phase $S$ acts as a velocity potential: ${\bf v} = (\hbar / m) \nabla
S$. By choosing  $S=\kappa \phi$, where $\phi$ is the angle around the
 $z$-axis and $\kappa$ is an integer, one has vortex states
with  tangential velocity
\begin{equation}
 v =    {\hbar \over m r_{\perp} } \kappa  \; ,
\label{eq:v}
\end{equation}
with $r_{\perp}^2 = x^2 + y^2$.
The number $\kappa$ is the quantum of circulation and the angular
momentum along $z$ is $N \kappa \hbar$.   Now, one has to put  the
complex wave function  $\Psi$ in place of $\psi$ in the energy
functional (\ref{eq:e}). Using the adimensional quantities defined
in Eqs.~(\ref{eq:r1}, \ref{eq:e1}, \ref{eq:psi1}) and $r_{1\perp}^2 =
x_1^2 + y_1^2$, the resulting  Gross-Pitaevskii functional becomes
\begin{equation}
{ E_1 \over N } = \int\! d{\bf r}_1  \left[ |\nabla_1 \psi_1 ({\bf
r}_1)|^2 + (\kappa^2 r_{1\perp}^{-2}+r_{1\perp}^2+ \lambda^2 z_1^2)
|\psi_1 ({\bf r}_1)|^2 + {u_1 \over 2}  |\psi_1 ({\bf r}_1)|^4 \right]
\; ,
\label{eq:vortexfunct}
\end{equation}
which  differs from functional (\ref{eq:functional}) because of the
centrifugal term. The corresponding nonlinear Schr\"odinger equation
is
\begin{equation}
\left[ -\nabla_1^2 + \kappa^2 r_{1\perp}^{-2}
+ r_{1\perp}^2 + \lambda^2 z_1^2 +  u_1 |\psi_1 ({\bf r}_1)|^2 \right]
\psi_1 ({\bf r}_1) =  2 \mu_1 \psi_1 ({\bf r}_1) \; .
\label{eq:vortexnlse}
\end{equation}
Due to the presence of the centrifugal term, the solution of this
equation  for $\kappa \ne 0$ has to vanish on the $z$-axis.

For noninteracting particles  one falls again in
the case of the stationary Schr\"odinger equation for the anisotropic
harmonic potential.   For instance the $\kappa=1$ solution has the
form
\begin{equation}
\psi_1 ({\bf r}_1) \propto  r_{1\perp} \exp \left[ - {1 \over 2}
( r_{1\perp}^2 + \lambda z_1^2) \right] \; .
\label{eq:vortexpsi1ho}
\end{equation}
The energy per particle of the $\kappa \ne 0$ states of the
anisotropic harmonic oscillator is simply $\kappa \hbar \omega_\perp$
plus the ground state energy. In our dimensionless notation: $ \mu_1
= 1 + (\lambda / 2) + \kappa$.

In the interacting case the kinetic energy can not be neglegted even
for large $N$, since it determines the structure of the vortex core.
In particular, the balance between the kinetic energy and the
interaction energy fixes a typical distance over which the condensate
wave function can heal. For a dilute Bose gas the {\it healing
length} is given by\cite{Pit61}
\begin{equation}
\xi = (8 \pi \rho a)^{-1/2}
\label{eq:csi}
\end{equation}
where $\rho$ is the density of the system. In the case of a vortex
it corresponds to  the distance over which the wave function
increases from zero, on the vortex axis, to the bulk density. For the
trapped atoms in the $N \to \infty$ limit we have seen that the
central density of the cloud is about $\sqrt{2 \mu_1 / u_1}$, where
$u_1$ and $\mu_1$ are given in Eqs.~(\ref{eq:u1}) and (\ref{eq:mu1}),
so that the healing length (in units $a_\perp$)  is  $ \xi_1 \simeq
(2 \mu_1)^{-1/2}$. Since the radius $R_1$ of the cloud, in the same
units,  is of the order of $(2 \mu_1)^{1/2}$, one has\cite{Bay95}
\begin{equation}
{\xi_1 \over R_1 }  = { 1 \over 2 \mu_1 } = { 1 \over R_1^2 } \; ,
\label{eq:xi1overr1}
\end{equation}
or, equivalently,
\begin{equation}
{\xi \over R }  =  \left( { a_\perp \over R } \right)^2 \; .
\label{eq:xioverr}
\end{equation}
Thus the healing length is small compared with the size of the cloud
if  $R$ is much bigger than $a_\perp$.

Vortex states play an important role in characterizing the superfluid
properties of Bose  systems, as is well known in the case of
superfluid Helium\cite{Don91}. The critical  angular velocity required
to produce vortex states is easily calculated once the energies of the
states with and without vortices is known. One has to compare the
energy of a vortex state in frame rotating with angular frequency
$\Omega$, that is $(E-\Omega L_z)$,  with the energy of the ground
state with no vortices.  Since the angular momentum per particle is
$\kappa  \hbar$, the critical angular velocity, in $\omega_\perp$
units, is
\begin{equation}
\Omega_c = \kappa^{-1} [(E_1/N)_{\kappa} - (E_1/N)_0]   \; .
\label{eq:omegac}
\end{equation}
In the noninteracting case the difference of energy per particle is
simply $\kappa$, so that $\Omega_c=1$;  the critical  angular velocity
is just the angular  frequency of the trap in the $x,y$ plane.

\section{Numerical procedure}
\label{section3}

The main purpose of this work is the numerical minimization of the
Gross-Pitaevskii functional (\ref{eq:vortexfunct}) in order to
calculate the properties of the ground state ($\kappa=0$) and of
vortex states  ($\kappa \ne 0$) for given values of the parameters $N,
\lambda, a_\perp$ and $a$. A method of direct minimization is provided
by the steepest descent approach. In brief, it consists of projecting
onto the minimum of the functional an initial trial  state by
propagating it in imaginary time. A time dependent wave function
$\psi_1({\bf r}_1,t)$, where $t$ is a fictitious time variable, is
evaluated at different time steps, starting from an arbitrary trial
function and converging to the exact solution  $\psi_1({\bf
r}_1,\infty) \equiv \psi_1({\bf r}_1)$. The time evolution can be
formulated in terms of the equation
\begin{equation}
{ \partial \over \partial t } \psi_1({\bf r}_1,t) =
- { \bar \delta E_1/N \over \bar \delta \psi_1({\bf r}_1,t) } \; ,
\label{eq:psidot}
\end{equation}
where $\bar \delta$ indicates the constrained functional derivative
that preserves the normalization.  This equation defines a
trajectory in the wave function space (and the fictitious time is
just a label for different configurations) in which at each step one
moves a little bit down the gradient $-  \bar \delta E/ \bar  \delta
\psi$. The constrained functional derivative is obtained by adding
the normalization condition to the functional derivative
\begin{equation}
{  \delta E_1/N \over  \delta \psi_1({\bf r}_1,t) } =
H \psi_1({\bf r}_1,t)
\; ,
\label{eq:hpsi}
\end{equation}
where $H$ depends nonlinearly on $\psi_1$. The end
product is the self-consistent minimization of the energy, which
corresponds to $ (\partial \psi_1/  \partial t)  =0$ or, including the
normalization, to the equation $H \psi_1 = 2 \mu_1 \psi_1$, which
coincides with the nonlinear Schr\"odinger equation
(\ref{eq:vortexnlse}).  In practice one chooses an arbitrary time step
$\Delta t$ and iterates the equation
\begin{equation}
\psi_1({\bf r}_1,t+\Delta t) \simeq \psi_1({\bf r}_1,t)
- \Delta t\  H \psi_1({\bf r}_1,t)  \; ,
\label{eq:iteration}
\end{equation}
by normalizing $\psi_1$ to $1$ at each iteration. The time  step
$\Delta t$ controls the rate of convergence. Several methods have
been proposed in the recent literature (see for instance
Ref.~\cite{Sti89} and reference therein) to improve the steepest
descent method of functional minimization; however, the
Gross-Pitaevskii functional is much simpler than the typical
functionals used in strongly correlated systems and the steepest
descent method described above is enough efficient for our purposes.

In practice one has to discretize the $(r_{1\perp},z)$-space with a
two-dimensional grid of points, so that the wave function becomes a
matrix. At each time step the matrix elements are changed as in
Eq.~(\ref{eq:iteration}), where the derivatives entering the
Hamiltonian are evaluated by means of finite difference formulae. The
algorithm  can be tested by comparing the results of the
noninteracting case with the analytical solution of the anisotropic
harmonic  potential. In the interacting case, with large $N$, it is
also possible to compare the numerical results with the analytic
solution (\ref{eq:thomasfermi}). Another test of accuracy is given
by the virial theorem, which fixes rigorous relationships among the
different contributions to the kinetic and potential energy of the
system at any value of $N$.

The system is sufficiently well described using a grid of $50 \times
50$ points in the range  $0 < r_{1\perp} < 5$, and the same for $z$.
The number of iterations in imaginary time depends on the degree of
convergence required and the goodness of the initial trial wave
function. The latter can be one of the two analytical limits already
discussed, but the final results do not depend on the trial wave
function. Typically we use $2000 \div 10000$ iterations. Since the
internal energy is a local functional, each iteration is very fast,
so that the functional minimization takes no more than $2 \div 3$
minutes of CPU on a DEC-Alpha processor.

\section{Results}
\label{section4}

\subsection{Positive scattering length: $^{87}$Rb  }

As an example of atoms with repulsive interaction we choose $^{87}$Rb,
as  in the experiment of Ref.~\cite{And95}. The $s$-wave triplet-spin
scattering length is in the range $85a_0 < a < 140a_0$, where $a_0$ is
the Bohr radius\cite{Gar95}. In our analysis we use $a=100a_0$. The
asymmetry parameter of the experimental trap is $\lambda =
\omega_z/\omega_\perp= \sqrt{8}$. The axial frequency $\omega_z/2\pi$
is taken to be $220$ Hz\cite{Cor95}. The corresponding characteristic
length is $a_\perp= 1.222 \times 10^{-4}$ cm and the ratio between the
scattering and the oscillator lengths is $a/a_\perp=6.47 \times
10^{-3}$.

We minimize the Gross-Pitaevskii functional in a wide range of
particle number $N$. Results for the chemical potential and  the
energy per particle are shown in Table~\ref{table1}. Both quantities
are expressed in units of $\hbar \omega_\perp$, or of the equivalent
temperature $\hbar \omega_\perp/k_B = 3.73$ nK. The partial
contributions to the energy per particle coming from the kinetic
energy ($kin$), the harmonic oscillator potential ($ho$) and the
internal potential energy ($pot$) are also given. The $N=1$ case
coincides with the noninteracting anisotropic harmonic oscillator: in
this case the internal potential energy vanishes, the kinetic energy
and the harmonic oscillator potential energy are equal, the chemical
potential and the total energy per particle are both equal to the
analytic value  $(1+\lambda/2)=2.41$.  When $N$ increases the
repulsion among atoms  tends to lower the central density, expanding
the cloud of atoms towards regions  where the trapping potential is
higher. This produces an increase  of both the internal and the
harmonic oscillator potential energy per particle. Conversely, the
kinetic energy per particle decreases because the density
distribution is flattened. In the strongly repulsive limit, $N \to
\infty$, one should find that the internal potential energy is much
greater than the kinetic energy, which is the case discussed in
Sec.~\ref{subsectionb}.  Indeed the convergence towards this limit
turns out to be rather slow. An approximate estimate of the kinetic
energy per particle can be obtained assuming the wave function to be
a gaussian, having a width of the order of the radius $R$ of the
cloud\cite{Bay95}. In this model the kinetic energy is of the order
of $\hbar^2/(2mR^2)$, which is much smaller than the internal
potential energy even for relatively small $N$ (a factor $10^{-2}$ is
reported in Ref.~\cite{Bay95} for $N=2000$).  The discrepancy between
the gaussian approximation and the exact solution is well understood
by looking at the effect of the surface structure of the cloud. In
Fig.~\ref{fig1}  we plot the profiles of the wave function along the
$x$- and $z$-axis for several values of $N$.  The noninteracting case
is shown as a dashed line. Increasing $N$ the central  density is
significantly lowered. The density in the cloud becomes almost flat
and it is well approximated  by the analytic solution
(\ref{eq:thomasfermi}) valid in the strongly repulsive limit. At the
surface the wave function vanishes gradually, the typical decay
length being almost independent of $N$. The contribution of the
surface to the kinetic energy remains sizable even for large $N$, so
that the kinetic energy is larger than the gaussian estimate
$\hbar^2/(2mR^2)$.  A typical profile of the condensate wave function
$\psi_1$ is plotted along the  $x$-axis for $N=5000$ in
Fig.~\ref{fig2}. The exact minimization of the Gross-Pitaevskii
functional (solid line) is compared with the noninteracting case
(dashed line) and the strongly repulsive limit (dot-dashed).

Simple relationships among the different contributions to the total
energy  are obtained by means of the virial theorem . When
applied to the anisotropic trap it gives the rigorous relation
\begin{equation}
{ \langle p_x^2 \rangle \over 2m } - {m \over 2} \omega_\perp
\langle x^2 \rangle  + {1 \over 2}  E_{pot} = 0
\label{eq:virial}
\end{equation}
and analogously for $y$ and $z$. Summing over the three equations for
$x$, $y$, and $z$ one finds
\begin{equation}
2 E_{kin} - 2 E_{ho} + 3 E_{pot} = 0  \; .
\label{eq:totalvirial}
\end{equation}
One can easily see that the numerical results in Table~\ref{table1}
agree very well with this relation.

The average  size of the cloud in both directions  can be easily
evaluated once the ground state wave function is known.  In the last
two columns of Table~\ref{table1} we report the quantities
$\sqrt{\langle x_1^2 \rangle}$ and $\sqrt{\langle z_1^2 \rangle}$.
When $N$ increases the quantity   $\langle x_1^2 \rangle$ deviates
rapidly from the noninteracting value  $1/2$, reflecting the
spreading of the atom distribution in the direction of the softer
trapping potential. The increase of $\langle z_1^2 \rangle$ is
slower,  but  never negligible. On can compare the results of the
numerical solution of the Gross-Pitaevskii equations with the ones
obtained in the strongly repulsive limit (see  Eq.~(\ref{eq:z1ave})).
In Table~\ref{table2} we give the approximated  chemical potential
(\ref{eq:mu1})  and the average sizes (\ref{eq:z1ave}),  using the
same input parameters (frequencies of the trap and scattering
length). Comparing these values with the ones in Table~\ref{table1},
one clearly sees that the strongly repulsive limit provides good
estimates for the  quantities $\langle x_1^2 \rangle$ and  $\langle
z_1^2  \rangle$. This means that the behavior of the surface
structure, which is  very  different in the exact and approximated
wave functions, does not  affect significantly the average sizes of
the cloud. Actually, the estimate of  $\langle x_1^2 \rangle$ is
better than the one for $\langle z_1^2 \rangle$, since the exact wave
function approaches more rapidly the  one of the strongly repulsive
limit  in the direction of the softer  trapping potential. The
approximated values of the chemical potential are close to the exact
ones  for $N$ very large. The quality of the strongly repulsive
approximation is improved in systems with greater values of the parameter
$u_1$, as in the case of the sodium vapor used in the experiment of
Ref.~\cite{Dav95}, where $N \simeq 10^{5}$ and $u_1$ is of the order
of $10^3$.

Another interesting quantity which  can be easily calculated from the
ground state wave function is the aspect ratio of the velocity
distribution, that is the ratio $\sqrt{ \langle p^2_z \rangle /
\langle p^2_x \rangle }$. This quantity is equal to $\sqrt{\lambda}$
in the noninteracting case and should approach $\lambda$ in the
strongly repulsive limit. The numerical results, as a function of
$N$, are shown in Fig.~\ref{fig3}. The two limiting cases are shown
as dashed lines. One clearly sees that the convergence to the value
$2.828=\lambda$ is very slow; the aspect ratio remains well below the
asymptotic value even for $N=20000$.  The aspect ratio measured in
Ref.~\cite{And95} is estimated to be about $50$ \% larger than the
noninteracting value, while the number of particles is  of the order
of $5000$. The agreement with our results is  good, even  if one has
to consider that the experimental estimate implicitly assumes a
ballistic expansion of the atoms after switching off the external
trap. The effects of the interaction on the expansion of the gas
should be explicitly taken into account in order to draw more
definitive conclusions.

Let us now consider the vortex states. In Fig.~\ref{fig4} we  show the
wave function of a cloud of $5000$ atoms; the $\kappa=1$  wave
function (b), which corresponds to atoms flowing around the $z$-axis
with angular momentum $N \hbar$,   is compared with the $\kappa=0$
ground state (a).  The atoms are pushed away  from the axis forming a
toroidal cloud. From the energy of the vortex states  we calculate
the critical angular velocity, through Eq.~(\ref{eq:omegac}). The
results for $\kappa=1$ are shown in Fig.~\ref{fig5}. The critical
angular velocity decreases rapidly with $N$. For $N>5000$ it is less
than $40$\% of the noninteracting value, given by the transverse
angular frequency $\omega_\perp$ of the trap.   A rough estimate of
the critical frequency in the large $N$ limit  is given by\cite{Bay95}
$\Omega_c/\omega_\perp \simeq (a_\perp/R)^2  \ln(R/\xi)$, where  $R$
is the  radius of the cloud. The {\it healing length}  is
the distance over which the wave function grows from zero to the  {\it
bulk} value. In the limit of large systems it  can
be approximated by Eq.~(\ref{eq:csi}) with $\rho$ equal to
the  density in the central part of the toroidal distribution.  Both
the estimate of $\xi$ and $\Omega_c$ obtained in this way are in
qualitative agreement  with the behavior of  the  numerical
solutions.   One can also find solutions for $\kappa >1$. The critical
velocity turns out to increase with $\kappa$. For instance, the
critical frequency $\Omega_c/2\pi$ for the creation of vortices in a
system of  $10000$ atoms is $26$, $35$ and $41$ Hz for $\kappa=1,2$
and  $3$, respectively.

Finally it is worth recalling that  the dimensionless parameter
characterizing the effects of the interactions in the Gross-Pitaevskii
equations is given by $u_1 = 8 \pi a N / a_\perp $ (see
Eqs.~(\ref{eq:u1},\ref{eq:functional})).  This implies that all the
results obtained in the present work can be applied, with a proper
rescaling of the variable $N$, to different choices for $a$ and/or
$a_\perp$.  For instance, changing the axial  frequency  of the trap
from $220$ Hz to $120$ Hz, so that $a_\perp$  increases by a factor
$\sqrt{11/6}$, is equivalent to keeping $a$ and $a_\perp$
unchanged and  reducing the number of atoms by the same factor
$\sqrt{11/6}$.

\subsection{Negative scattering length: $^7$Li }

As an example of atoms with attractive interaction  we choose $^7$Li,
as in the experiment of Ref.~\cite{Bra95}. The $s$-wave triplet-spin
scattering length is  $-27a_0$\cite{Abr95}.  The axial frequency
reported in Ref.~\cite{Bra95} is $\omega_z/2\pi=117$ Hz and the
corresponding characteristic length is $a_\perp= 2.972 \times 10^{-4}$
cm, thereby yielding a ratio $|a|/a_\perp=0.48 \times 10^{-3}$.  The
transverse frequency is  $\omega_z/2\pi=163$ Hz, so that the
asymmetry parameter is $\lambda = \omega_z/\omega_\perp= 0.72$.

The first important point to stress is that Gross-Pitaevskii
functional has no global minimum for negative scattering length. This
reflects the tendency of the system to collapse. For spatially
inhomogeneous systems, however, the zero-point energy can exceed the
attractive potential, producing local minima of the functional when
the density of atoms is not too high. The  nonlinear stationary
Schr\"odinger equation provides the solutions $\psi$ for which $\delta
E / \delta \psi =0$, but does not say anything about the stability of
these solutions. A proper treatment of the stability requires a time
dependent theory\cite{Rup95,Fet95}. The minimization of the
Gross-Pitaevskii functional with the steepest descent method explores
the configuration space with axial symmetry near the local minimum.

In Fig.~\ref{fig6}  we show the results for the wave function along
the $x$- and $z$-axis for several values of $N$. As in Fig.~\ref{fig1}
we plot the  noninteracting case with a dashed line. Here the  vapor
extends more along $z$ than along $x$, just because the external
potential of the trap is softer in the axial  direction ($\lambda <
1$). Apart from this purely  geometrical fact, the most striking
difference with respect to the repulsive case is that here the
central density of the cloud increases rapidly with $N$. This is  the
effect of adding more and more attractive potential energy.  When the
central density reaches a certain critical limit the system
collapses. In term  of functional minimization this implies that the
convergence towards the local minimum becomes slower and slower,
until a critical $N$ above which the energy falls down and does not
converge anymore.  In $^7$Li, with the input parameters given above,
the critical number $N$ turns out to be about $1400$.

Looking at  Fig.~\ref{fig6} one also notices  that the wave function
changes its form  in the same way along $x$ and $z$.  Both the average
sizes $\sqrt{\langle x^2 \rangle}$ and $\sqrt{\langle z^2 \rangle}$
decreases slowly  when $N$ increases. For instance, for $N=1000$ one has
$\sqrt{\langle x_1^2 \rangle}=0.62$ and $\sqrt{\langle z_1^2
\rangle}=0.69$, both values being about $15$\% smaller than the ones
in the noninteracting case.  The ratio
$\sqrt{ \langle x^2 \rangle / \langle z^2 \rangle }$ is
practically independent of $N$. For instance for $N=1$ it is equal to
$\sqrt{\lambda}= 0.85$, while for  $N=1000$ it is $0.90$, with an
increase of only $5$\%. The aspect ratio of the velocity distribution,
which is equal to $\sqrt{\lambda}$ in the noninteracting case, behaves
in the same way. Even the energy per particle depends smoothly on $N$.
In units $\hbar \omega_\perp$, it is equal to $1.36$ and $1.15$ for
$N=1$ and $N=1000$, respectively.

Coming back to the question of the stability, we notice that, when
the local minimum associated with wave functions of the form shown in
Fig.~\ref{fig6} disappears, nothing prevents {\it a priori} the
existence of other local minima associated with different
configurations. Such configurations should have local density lower
than the critical one. A natural way to obtain a favorable situation
is to move the atoms away from the $z$-axis,  conserving the total
number of particles. This happens in the presence of a   vortex. In
Fig.~\ref{fig7} we show the wave function for $1000$\ $^7$Li  atoms
with no vortices (a) and with an axial vortex of unit circulation (b).
We use the same  units in both cases, so one can see that the
maximum value of the wave function inside the toroidal distribution of
the vortex is approximately a factor two lower than the central value
in the state with no vorticity (the density is  four times smaller).
The critical angular frequency for the formation of the vortex  state
in Fig.~\ref{fig7} is $1.12$ times the transverse  angular frequency
of the trap. In systems with  attractive interaction the critical
angular velocity is larger than for noninteracting particles, while
the opposite is true for repulsive interaction. This is because it
costs internal potential energy to lower the average density, as the
vortex does, for attractive interactions.   However, once a vortex is
created, the corresponding state is more stable than in the absence of
vorticity: one can put more atoms inside the rotating cloud before
reaching the critical density for the final collapse. Indeed we find
local minima of the Gross-Pitaevskii functional for $N$ much larger
than $1400$ if $\kappa > 0$. We show three examples in
Fig.~\ref{fig8}. One notices that the maximum  density of the
$\kappa=1$ state slightly increases from the case $N=1000$
(Fig.~\ref{fig7}b) to the case  $N=3500$ (top of Fig.~\ref{fig8}),
however remaining well below the value of the central density in the
state without vorticity (Fig.~\ref{fig7}a). The three vortices in
Fig.~\ref{fig8} have almost the same peak density, but very different
number of particles. They correspond to well defined local minima of
the functional. If the number of particles is increased, one finds
again critical values of $N$ for which the minima disappear.  For
$\kappa=1$ we find a critical value of $N \simeq 4000$; for
$\kappa=2$ and $3$ we find critical values of $6500$ and $8300$,
respectively. It is worth mentioning that the number of particles in
the condensate reported in the experimental work of Ref.~\cite{Bra95}
is  an order of magnitude  higher than the critical value for the
stability of the Gross-Pitaevskii solution without vorticity ($N
\simeq 1400$). This discrepancy between the experimental finding of
Ref.~\cite{Bra95} and the predictions of the Gross-Pitaevskii theory
could be significantly reduced if one assumes the existence of a
vortex in the atomic cloud.

\section{Conclusions}
\label{conclusion}

In this paper we have solved the Gross-Pitaevskii equations  for a
dilute gas of alkali atoms in anisotropic magnetic traps by numerical
minimization of the total energy.  The theory provides the condensate
wave function at $T=0$ for states with and without vorticity. The
comparison with approximate models, such as the noninteracting gas and
the strongly repulsive limit, has been carefully explored. We have
explicitly discussed the results for $^{87}$Rb (positive scattering
length)  and $^7$Li (negative scattering length) since these elements
have been recently used in the first successful measurements of Bose
Einstein condensation\cite{And95,Bra95}.  We summarize here the main
results of the present analysis.

\begin{itemize}

\item We have explored in a systematic way the density distribution
and the energy systematics of the atomic clouds. The exact condensate
wave function is flat in the interior  and    vanishes
smoothly at the surface. The contribution of the surface to the
kinetic energy per particle remains sizable even for relatively large
$N$, differently from the predictions of approximated models recently
proposed. This affects significantly the  behavior of the aspect ratio
of the velocity distribution. In the case of positive scattering
length the aspect ratio is larger than the value $\sqrt{\lambda}$
given by the noninteracting model but smaller than the value $\lambda$
given by the strongly repulsive limit. The values calculated  for
$^{87}$Rb are in agreement with the  experimental findings of
Ref.~\cite{And95}.

\item We have studied the properties of vortex states.   For systems
with repulsive interaction the critical angular velocity for the
formation of vortices decreases rapidly with $N$ with respect to the
value of the noninteracting gas. Conversely, it increases with $N$ in
systems with attractive interaction. The most striking feature of
vortex states is the tendency to lower the peak density in the cloud
of atoms. This tendency has a dramatic effect for systems with
attractive interaction, where high values of the peak density can
produce the disappearance of the local minimum of the functional,
i.e, the collapse of the system. In $^7$Li this happens for $N \simeq
1400$.  It turns out that the presence of a vortex increases the
stability of the system, in the sense that local minima with larger
$N$ can be found.   We have shown the results up to $N=8000$ with
circulation number $\kappa=3$. Higher values of $N$ can be obtained
by increasing $\kappa$.

\end{itemize}

Further work is planned in order to study in more details the velocity
distribution of the atomic vapor. Time dependent calculations are also
feasible within the same theoretical scheme.

\begin{figure}
\caption{Ground state wave function for $^{87}$Rb along the  $x$-axis
(upper part) and along the $z$-axis (lower part). Distances are  in
units $a_\perp$ (see Eq.~(\protect\ref{eq:aperpaz})). The dashed line
is the noninteracting case; the solid lines corresponds to $N=100,
200, 500,  1000, 2000, 5000$ and $10000$, in descending order of
central density. }   \label{fig1}
\end{figure}

\begin{figure}
\caption{Ground state wave function for $5000$ atoms of $^{87}$Rb.
Dashed line: noninteracting case (see Sec.~\protect\ref{subsectiona}).
Dot-dashed line: strongly repulsive limit (see
Sec.~\protect\ref{subsectionb}). Solid line: exact solution of
Gross-Pitaevskii functional. } \label{fig2}
\end{figure}

\begin{figure}
\caption{Ratio of the axial to transverse average velocity as a
function  of $N$ in $^{87}$Rb. The lower and upper dashed lines
corresponds to  $\protect \sqrt{\lambda}$ and $\lambda$, respectively. }
\label{fig3}
\end{figure}

\begin{figure}
\caption{Wave function, in arbitrary units, of $5000$ \ $^{87}$Rb
atoms. a) Ground state. b) Vortex state with $\kappa=1$.  }
\label{fig4}
\end{figure}

\begin{figure}
\caption{Critical angular velocity, in units $\omega_\perp$,  for the
formation of $\kappa=1$ vortices in $^{87}$Rb vapor as a function
of $N$ } \label{fig5}
\end{figure}

\begin{figure}
\caption{Ground state wave function for $^7$Li along the $x$-axis
(upper part) and along the $z$-axis (lower part). Distances are in
units $a_\perp$ (see Eq.~(\protect\ref{eq:aperpaz})). The dashed line
is the noninteracting case; the solid lines corresponds to $N = 200,
500$ and  $1000$, in ascending order of central density. }
\label{fig6}
\end{figure}

\begin{figure}
\caption{Wave function, in arbitrary units, of $1000$ \ $^7$Li atoms.
a) Ground state. b) Vortex state with $\kappa=1$.  }
\label{fig7}
\end{figure}

\begin{figure}
\caption{Vortex state wave functions, in arbitrary units,  for
different values of $N$ and $\kappa$ in $^7$Li.  }
\label{fig8}
\end{figure}

\begin{table}

\caption{ Results for the ground state of $^{87}$Rb atoms in a trap
with $\protect \lambda=\protect \sqrt{8}$.   Chemical potential and
energy in  units $\hbar \omega_\perp$, with $2\pi  \omega_\perp=220$
Hz. Lengths in units $a_\perp$. }

\begin{tabular}{c|c|c|c|c|c|c|c}
N
&$\mu_1$&$(E_1/N)$&$(E_1/N)_{kin}$&$(E_1/N)_{ho}$&$(E_1/N)_{pot}$&
$\sqrt{\langle x_1^2 \rangle }$&$\sqrt{\langle z_1^2 \rangle }$\\
\tableline
1    &2.414&2.414&1.207&1.207&0.000&0.707&0.420\\
\tableline
100  &2.88&2.66&1.06&1.39&0.21&0.79&0.44\\
\tableline
200  &3.21&2.86&0.98&1.52&0.36&0.85&0.45\\
\tableline
500  &3.94&3.30&0.86&1.81&0.63&0.96&0.47\\
\tableline
1000 &4.77&3.84&0.76&2.15&0.93&1.08&0.50\\
\tableline
2000 &5.93&4.61&0.66&2.64&1.32&1.23&0.53\\
\tableline
5000 &8.14&6.12&0.54&3.57&2.02&1.47&0.59\\
\tableline
10000&10.5&7.76&0.45&4.57&2.74&1.69&0.65\\
\tableline
15000&12.2&8.98&0.41&5.31&3.26&1.84&0.70\\
\tableline
20000&13.7&9.98&0.38&5.91&3.68&1.94&0.73\\
\tableline
\end{tabular}
\label{table1}
\end{table}

\begin{table}

\caption{ Chemical potential (in units $\hbar \omega_\perp$) and
average transverse and vertical size (in units $a_\perp$) in the
strongly repulsive approximation  (Sec.~\protect\ref{subsectionb}),
for $^{87}$Rb in the same trap of Table~\protect\ref{table1}. }

\begin{tabular}{c|c|c|c}
N&$\mu_1$&$\sqrt{\langle x_1^2 \rangle }$&$\sqrt{\langle z_1^2 \rangle }$\\
\tableline
100  &1.60&0.68&0.24\\
\tableline
500  &3.05&0.94&0.33\\
\tableline
1000 &4.02&1.07&0.38\\
\tableline
5000 &7.66&1.48&0.52\\
\tableline
10000&10.1&1.70&0.60\\
\tableline
20000&13.3&1.94&0.69\\
\tableline
\end{tabular}
\label{table2}
\end{table}


\begin{references}

\bibitem{And95} M.H. Anderson, J. R. Ensher, M.R. Matthews, C.E.
Wieman, and E.A.  Cornell, Science {\bf 269}, 198 (1995).

\bibitem{Bra95} C.C. Bradley, C.A. Sackett, J.J. Tollett, and R.G.
Hulet, Phys. Rev. Lett. {\bf 75}, 1687 (1995).

\bibitem{Dav95} K.B. Davis, M.-O. Mewes, N.J. van Druten, D.S. Durfee,
D.M. Kurn, and W. Ketterle, preprint (1995)

\bibitem{Gri95} A. Griffin, D.W. Snoke, and S. Stringari,eds, {\it Bose
Einstein Condensation} (Cambridge Univ. Press, Cambridge, 1995)

\bibitem{Pit61} L.P. Pitaevskii, Zh. Eksp. Teor. Fiz. {\bf 40}, 646
(1961)  [Sov. Phys. JETP {\bf 13}, 451 (1961)]; E.P. Gross, Nuovo
Cimento {\bf 20}, 454 (1961); E.P. Gross, J. Math. Phys. {\bf 4}, 195
(1963)

\bibitem{Edw95} M. Edwards and K. Burnett, Phys. Rev. A {\bf 51}, 1382
(1995)

\bibitem{Rup95} P.A.  Ruprecht, M.J. Holland, K. Burnett, and M.
Edwards, Phys.  Rev.  A {\bf 51}, 4704 (1995).

\bibitem{Fet95} A. L. Fetter, preprint, cond-mat/9510037 (1995)

\bibitem{Bay95} G. Baym and C. Pethick, preprint, cond-mat/9508040
(1995)

\bibitem{Str95} S. Stringari, preprint, cond-mat/9509166 (1995)

\bibitem{Don91} R. J. Donnelly, {\it Quantized Vortices in Helium II},
Cambridge Studies in Low Temperature Physics (Cambridge University
Press, Cambridge, England, 1991)

\bibitem{Sti89} I.  \v{S}tich, R. Car, M. Parrinello, and S. Baroni,
Phys. Rev. B {\bf 39}, 4997 (1989)

\bibitem{Gar95} J.R. Gardner, R.A. Cline, J.D. Miller, D.J. Heinzen,
H.M.J.M. Boesten, and B.J. Verhaar, Phys. Rev. Lett. {\bf 74}, 3764
(1995)

\bibitem{Cor95} The value used here, which is  different from the  one
reported in Ref.~\protect\cite{And95} ($120$ Hz),  is expected to
describe better the  same experimental trap   (E.A. Cornell, private
communication).

\bibitem{Abr95} E.R.I. Abraham, W.I. McAlexander, C.A. Sackett,  and
R.G. Hulet, Phys. Rev. Lett {\bf 74}, 1315 (1995)

\end{references}
\end{document}